\documentstyle[prb,aps,12pt]{revtex}
\begin{document}
\draft
\title{Density of Neutral Solitons in Weakly Disordered Peierls
Chains}
\author{M.~V.~Mostovoy\cite{Perm}, M.~T.~Figge and J.~Knoester}
\address{Institute for Theoretical Physics, Materials Science Center\\
University of Groningen, Nijenborgh 4, 9747 AG Groningen, The Netherlands}
\date{\today}
\maketitle

\begin{abstract}
\widetext
\leftskip 54.8pt
\rightskip 54.8pt

We study the effects of weak off-diagonal disorder on Peierls
systems with a doubly degenerate ground state.  We show that for
these systems disorder in the electron hopping amplitudes induces
a finite density of solitons in the minimal-energy lattice
configuration of a single chain.  These disorder-induced
dimerization kinks are neutral and have spin $\frac12$.  Using a
continuum model for the Peierls chain and treating the lattice
classically, we analytically calculate the average free energy
and density of kinks.  We compare these results to numerical
calculations for a discrete model and discuss the implications of
the kinks for the optical and magnetic properties of the
conjugated polymer {\em trans}-polyacetylene.

\end{abstract}


\section{Introduction}

Recently we considered the effects of weak disorder in the
electron hopping amplitudes on the lattice configuration of
Peierls systems with a doubly degenerate ground state, using the
conjugated polymer {\em trans}-polyacetylene as an
example.\cite{MFK} In the absence of disorder the ground state of
a {\em trans}-polyacetylene chain is uniformly dimerized due to
the interaction of the lattice with the half-filled band of
$\pi$-electrons propagating along the chain.\cite{SSH} The ground
state energy is independent of the sign of the dimerization, {\em
i.e.}, it is equal for the two carbon-carbon bond alternations
$\ldots$-long-short-long-short-$\ldots$ and
$\ldots$-short-long-short-long-$\ldots$.  Disorder in the
electron hopping amplitudes, originating, {\em e.g.}, from random
twists of bonds, removes this degeneracy.  This can be understood
from the fact that such conformational disorder reduces the
overlap between the electron orbitals of neighboring carbon atoms
and thus increases the chain energy.  The energy increase is less
for the long bonds (with relatively small hopping amplitudes)
than for the stronger short bonds.

Due to the random nature of the disorder fluctuations, the
preferable sign of the dimerization varies along the chain.  This
explains why domain walls (kinks), separating regions with
positive and negative dimerization, can be stabilized by
disorder.\cite{MFK,Iw,Wa} While in the absence of disorder, kinks
(or solitons) are topological excitations with a rather high
energy ($\sim0.5$eV in the case of {\em trans}-polyacetylene),
they do appear in the minimal-energy lattice configuration of a
disordered chain.  This was first noted in numerical simulations
of the Su-Schrieffer-Heerger (SSH) model.\cite{Iw,Wa} The large
kink energy is compensated by adjusting the sign of the chain
dimerization to the disorder fluctuations in the intervals
between the kinks.  The weaker is the disorder, the longer the
distance between neighboring domain walls has to be.  Simple
arguments, based on an estimate of the size of the typical
disorder fluctuation that stabilizes a kink-antikink pair, show
that at weak disorder the average density of disorder-induced
kinks in the minimal-energy lattice configuration is proportional
to the strength of the disorder.\cite{MFK}

The effect of off-diagonal disorder on a Peierls chain is similar
to the effect of nonzero temperature.  It is well-known, that the
$Z_2$ symmetry in one-dimensional systems can be spontaneously
broken only at $T=0$.  At any non-zero temperature the symmetry
is restored by thermally induced kinks.  In this case the kink
creation energy is compensated by its large entropy, as kinks can
be located at any place in the chain.  The topological nature of
kinks is responsible for the destruction of the long range order
in isolated Peierls chains, both at arbitrarily small temperature
and at arbitrarily weak disorder.

In this paper we put a firm basis under the results which we
obtained through simple scaling arguments in Ref.[\CITE{MFK}], by
giving a detailed calculation of the free energy and the density
of kinks for half-filled Peierls chains with off-diagonal
disorder.  The continuum model describing such chains is
introduced in Sec.~\ref{model}.  In Sec.~\ref{stabil} we briefly
repeat the arguments of Ref.[\CITE{MFK}], showing that the energy
of a disordered chain can be decreased by creating a
kink-antikink pair.  This implies a special role of kinks.  In
Sec.~\ref{average} we consider the partition function of a
disordered chain, treating the lattice classically.  Integrating
out small lattice fluctuations, we obtain an effective free
energy describing kinks.  The details of this integration can be
found in the Appendix.  We then use the transfer matrix approach
to reduce the averaging of the chain's free energy over the
disorder realizations to the averaging of the wave function that
describes the relaxation of a spin $1/2$ in a magnetic field with
one random component.  For long chains, the latter average can be
calculated analytically.  The derivative of the thus obtained
free energy with respect to the chemical potential of kinks gives
the average density of kinks, induced both thermally and by
disorder (Sec.~\ref{density}).

As we shall see, in the continuum model with white Gaussian
disorder and a classical lattice, the entropy of kinks becomes
negative below a certain temperature $T_0$ which depends on the
disorder strength.  Therefore, in Sec.~\ref{num} we also study
the generation of kinks by disorder in a discrete model, which
does not suffer from this pathology.  The discrete model is the
one-dimensional Random Field Ising Model (RFIM).  It was realized
long ago by Imry and Ma,\cite{IM} that kinks destroy the long
range order in this model even at zero temperature.  We present
results of numerical simulations of the average density of kinks
in the RFIM as a function of both the disorder strength and
temperature and compare to the analytical results for the
continuum model.  In Sec.~\ref{dis} we summarize and conclude.
We also connect to previous work and discuss the effects of
disorder-induced kinks on the optical and magnetic properties of
quasi-one-dimensional Peierls systems.

\section{Continuum model of a disordered Peierls
chain}\label{model}

We start by considering a tight binding model that describes the
hopping of electrons along a chain of atoms.  The electron
hopping amplitudes depend on the interatomic distances and the
relative orientation of the electronic orbitals on neighboring
atoms.  Therefore, the hopping amplitudes are affected by both
the lattice motion (the displacement of the atoms parallel to the
chain) and conformational disorder (chain twists).   Let
$t_0$ denote the hopping amplitude between neighboring atoms in
a perfect rigid chain of equidistant atoms with lattice constant
$a$.  Then, in the presence of atomic displacements and
conformational disorder, the hopping amplitudes may be written,
\begin{equation} \label{hop} t_{m,m+1} = t_0 + \alpha (u_m -
u_{m+1}) + \delta t_{m,m+1}\;.
\end{equation}
Here, the second term is the SSH-type of electron-phonon
interaction,\cite{SSH} with the coupling constant $\alpha$ and
$u_m$ the displacement of the $m$'th atom from its
uniform-lattice position.  The third term is a random
contribution resulting from the conformational disorder.  While
the lattice displacements $u_m$ are dynamic variables, we will
assume that the fluctuations $\delta t_{m,m+1}$ are frozen
(``quenched'' disorder).

The Peierls order parameter is the alternating part of the
hopping amplitudes,
\begin{equation}
\Delta(2ma) = t_{2m-1,2m} - t_{2m,2m+1}\;,
\end{equation}
which consists of two parts,
\begin{equation}
\label{odpard}
\Delta(2ma) = \Delta_{lat}(2ma) + \eta(2ma)\;.
\end{equation}
The first part is the lattice dimerization,
\begin{equation}
\Delta_{lat}(2ma) = \alpha (u_{2m-1}-2u_{2m}+u_{2m+1})\;,
\label{Deltalat}
\end{equation}
which describes the alternating part of the hopping amplitude
determined by the shifts $u_m$ of the atoms and is the usual
order parameter of the SSH model.  The second term in
Eq.(\ref{odpard}) describes the disorder,
\begin{equation}
\eta(2ma) = \delta t_{2m-1,2m} - \delta t_{2m,2m+1}\;.
\end{equation}
We assume the random variations of the hopping amplitudes $\delta
t$ on different bonds to be independent.

For weak electron-phonon coupling and small disorder we can, in
analogy to Ref.[\CITE{TLM}], use a continuum description of both
electrons and lattice with the order parameter
\begin{equation}
\label{odpar}
\Delta(x) = \Delta_{lat}(x) + \eta(x)\;.
\end{equation}
Here, $\Delta_{lat}(x)$ is the continuum analog of
Eq.(\ref{Deltalat}) and $\eta(x)$ is the white noise disorder:
\begin{equation}
\label{Gauss}
\langle \eta (x) \eta(y) \rangle = A \delta(x - y) \;.
\end{equation}
Note, that while the random chain twists always decrease the
hopping amplitudes ($\delta t_{m,m+1} < 0$), $\eta(x)$ can be
both positive and negative, as it is the alternating part of the
fluctuations.

The Hamiltonian of the continuum model has the form:
\begin{equation}
H[\Delta_{lat}(x),\eta(x)] =
E_{lat}[\Delta_{lat}(x)] + H_{el}[\Delta(x)]\;\;.
\end{equation}
The first term is the harmonic lattice energy,
\begin{equation}
E_{lat}[\Delta_{lat}(x)] =
\frac{1}{\pi \lambda v_F} \int\!\!dx \Delta_{lat}(x)^2\;\;,
\end{equation}
where $\lambda = \frac{4 \alpha^2}{\pi t_0 K}$ is the
dimensionless electron-phonon coupling constant ($K$ is the
spring constant), $v_F = 2 a t_0$ is the bare value of the Fermi
velocity, and we set $\hbar=1$ [cf.  Ref.[\CITE{TLM}]].  In this
paper we treat the lattice classically, {\em i.e.}, we disregard
the lattice kinetic energy, which is reasonable for chains of
sufficiently heavy atoms.  It should be noted that for {\em
trans}-polyacetylene, which consists of relatively light
CH-groups, quantum lattice effects may be rather
important.\cite{Su82,Fradkin,Auerbach1,MK}

The electrons in the continuum model are described by,
\[
\psi_{\sigma}(x) =
\left(
 \begin{array}{c}
  \psi_{1 \sigma}(x) \\
  \psi_{2 \sigma}(x) \\
 \end{array}
\right) \;,
\]
where the two amplitudes $\psi_{1 \sigma}(x)$ and $\psi_{2
\sigma}(x)$ correspond to particles moving, respectively, to the
right and to the left with the (bare) Fermi velocity $v_F$, and
$\sigma$ is the spin projection.

The electron Hamiltonian has the form,
\begin{equation}
H_{el}[\Delta(x)] =
\sum_{\sigma = \pm1} \int\!\!dx \psi_{\sigma}^{\dagger}(x)
\left(
\frac{v_F}{i} \sigma_3 \frac{d}{dx} + \Delta(x) \sigma_1
\right)
\psi_{\sigma}(x) + H_{el-el}\;\;.
\end{equation}
The first term describes the motion of electrons in the presence
of both the chain distortion and the disorder ($\sigma_1$ and
$\sigma_3$ are the Pauli matrices), while the second term
describes the (Coulomb) interactions between electrons.  Apart
from the disorder and the electron-electron interaction term the
Hamiltonian of our model is the same as the Hamiltonian of the
continuum version of the SSH model.\cite{TLM}

In the absence of disorder the (half-filled) chain reaches its
minimal energy in either one of two uniformly dimerized
configurations $\Delta_{lat}(x) = \pm \Delta_0$.  This Peierls
instability was first found for noninteracting
electrons\cite{Peierls} ($H_{el-el}=0$).  It also occurs,
however, in the presence of electron-electron interactions, like,
{\em e.g.}, the on-site Hubbard repulsion $U$, which opens a gap
for charge excitations, but (in the absence of electron-phonon
interaction) leaves the spectrum of spin excitations
gapless.\cite{Horovitz,Krivnov,Voit} Numerical calculations have
shown that a moderately large $U$ can even increase the value of
the lattice dimerization.\cite{Hirsch,Mazumdar,Jeckelmann} As we
shall shortly see, the existence of two degenerate ground states,
which also implies the existence of kink solutions, is crucial
for the appearance of disorder-induced kinks.  The precise form
of $H_{el-el}$ does not affect this basic phenomenon; it only is
important to the extent that it determines the value of
$\Delta_0$ and the kink energy.  Therefore, in this paper we do
not specify the expression for $H_{el-el}$.

An important property of the dimerized Peierls state is the
existence of gaps in the spectra of spin and charge excitations.
For free electrons ($H_{el-el} = 0$) both gaps are equal, while
in the presence of Coulomb repulsion the spin gap is smaller than
the charge gap.\cite{Krivnov,Voit} In what follows, we will
assume the temperature to be much smaller than the spin gap, so
that we can neglect electronic excitations and replace
$H_{el}[\Delta(x)]$ by its ground state expectation value
(adiabatic approximation),
\begin{equation}
\label{adappr}
E_{el}[\Delta(x)] =
\langle 0 |H_{el}[\Delta(x)]| 0 \rangle\;.
\end{equation}

\section{Stabilization of kinks by off-diagonal
disorder}\label{stabil}

For the sake of completeness, we briefly repeat in this section
the arguments of Ref.[{\CITE{MFK}], which show that the
lowest-energy lattice configuration in the presence of disorder
may contain kinks.

At zero temperature the lattice configuration, {\em i.e.},
$\Delta_{lat}(x)$, has to be found by minimizing the total chain
energy,
\begin{equation}
E[\Delta_{lat}(x),\eta(x)] = E_{el}[\Delta(x)] +
E_{lat}[\Delta_{lat}(x)]\;,
\end{equation}
with respect to $\Delta_{lat}(x)$ at a given disorder realization
$\eta(x)$.  This makes $\Delta_{lat}(x)$ implicitly dependent on
$\eta(x)$.

As noted in Sec.~\ref{model}, in the absence of disorder the
total energy of a half-filled chain has two minima, $\Delta(x) =
\pm \Delta_0$, corresponding to two uniformly dimerized
configurations with the same energy.  Apart from the minima,
there are infinitely many lattice configurations that are nearly
perfect extrema of the total energy.  These are the multikink
configurations, in which a sequence of solitons and antisolitons
interpolate between $-\Delta_0$ and $+\Delta_0$ and vice versa
(Fig.~1).  A kink is locally stable, {\em i.e.}, the chain energy
increases, when its form is perturbed.  The energy of a multikink
configuration can be decreased only by changing the distances
between the kinks.  However, when the separation between
neighboring kinks is large compared to their size (which is of
the order of the correlation length $\xi_0 =
v_F/\Delta_0$),\cite{SSH} the change of the energy caused by
shifts of the kinks is exponentially small, so that the energy of
the configuration with $N$ kinks is approximately,
\begin{equation}
E_N = E_0 + N \mu\;.
\end{equation}
Here, $\mu$ is the energy needed to create a single kink (about
$0.5eV$ for {\em trans}-polyacetylene) and $E_0$ is the chain
energy in the absence of kinks.  The kinks can be either charged
and spinless, or neutral with spin $\frac{1}{2}$.\cite{SSH} In
the SSH model ($H_{el-el} = 0$), both types have the same
energy;\cite{SSH} if the Coulomb repulsion between electrons is
taken into account, the neutral soliton has lower
energy.\cite{KH,Cam} Since the on-site Coulomb repulsion in {\em
trans}-polyacetylene is appreciable,\cite{Kies} we will assume
that only neutral kinks are induced by disorder.

Next we find the change of the energy of a multikink
configuration due to weak disorder.  We will denote the lattice
configuration containing $N$ kinks, whose positions are described
by the $N$-dimensional vector $\mbox{\boldmath $z$} = (z_1,
z_2,\ldots,z_N)$, by $\Delta_N(x|\mbox{\boldmath $z$})$.
To first order in $\eta(x)$, the correction to the energy of the
configuration reads,
\begin{equation}
\label{EN}
\delta E_N = - \frac{2}{\pi \lambda v_F} \int \!\!dx
\Delta_N(x|\mbox{\boldmath $z$}) \eta(x) \;,
\end{equation}
where the extremum condition for the configuration
$\Delta_N(x|\mbox{\boldmath $z$})$ at zero disorder was used.

For instance, the change of the energy of the uniformly dimerized
configuration ($\Delta(x)=\Delta_0$) due to disorder is
\begin{equation}
\delta E_0 = -
\frac{2 \Delta_0}{\pi \lambda v_F} \int \!\!dx \eta(x) \;,
\end{equation}
while for the configuration with an antikink at $z_1$ and a
kink at $z_2$, such that the whole disorder fluctuation lies
between $z_1$ and $z_2$, the change of energy equals $\delta E_2
= - \delta E_0$, because between $z_1$ and $z_2$
$\,\Delta_2(x|z_1,z_2) \approx - \Delta_0$.  We thus see that in
the disordered chain, the configuration obtained by the
perturbation of a kink-antikink pair is energetically favorable
to the perturbed uniform configuration if
\begin{equation}
\label{pair}
- \int \!\!dx \eta(x) > \gamma_k \lambda v_F \;.
\end{equation}
Here we introduced $\gamma_k = \pi \mu / (2 \Delta_0)$, which
for free electrons (SSH model) equals 1.  Since the fluctuations
in $\int_0^L\!\!dx \eta(x)$ grow with the chain size $L$, for a
sufficiently long chain the inequality Eq.(\ref{pair}) can
certainly be fulfilled, no matter how small the disorder is.

We note that in addition to multikink configurations, also
non-topological lattice configurations exist that are extrema of
the total chain energy.  An example is the polaron.\cite{BK,CB}
However, since the polaron disturbs the lattice only locally, in
an interval of length $l \sim \xi_0$ (where $\xi_0$ is the
correlation length), a disorder fluctuation having large
magnitude in this interval is needed to stabilize it.  At weak
disorder the probability density of such a fluctuation is
exponentially small.  On the other hand, the disorder
fluctuations stabilizing kinks, {\em i.e.}, those that satisfy
Eq.(\ref{pair}), can have small magnitude compensated by a large
spatial extension given by the distance between neighboring
kinks.  The weight of such a fluctuation grows with its length
and becomes of the order unity if the length equals the average
distance between the disorder-induced kinks.\cite{MFK} In other
words, the fluctuations that stabilize kinks are not at all
suppressed.  This is why in what follows we take into account
only multikink configurations, disregarding all other large
variations of the Peierls order parameter.

\section{Free energy of disordered chains}\label{average}

We start by considering the partition function of a weakly
disordered Peierls chain.  As was mentioned above, we treat the
lattice classically and assume the temperature to be sufficiently
small to neglect electron excitations across the gap.  The
partition function is then given by the weighted sum over all
possible lattice configurations,
\begin{equation}
\label{parfun}
Z[\eta(x)] = \int D\Delta_{lat}
\exp \left\{ - \beta E[\Delta_{lat}(x),\eta(x)] \right\}\;,
\end{equation}
and is a functional of the disorder realization $\eta(x)$.
In Eq.(\ref{parfun}) we have omitted the part related to the
kinetic energy of the lattice, as ultimately we will only be
interested in the density of kinks in the chain.

It is, of course, impossible to perform the integration over
$\Delta_{lat}(x)$ in the partition function exactly, since even the
dependence of the electron energy on $\Delta(x)$ is, in general,
not known.  However, when the temperature tends to zero, only the
lattice configurations near the absolute minimum of the chain
energy contribute significantly to the partition function.  As we
discussed in the previous section, in the presence of weak
disorder the minimal energy lattice configuration is close to a
certain multikink configuration.  How many kinks it contains and
where they are located, depends on $\eta(x)$.  To make sure that
for any disorder realization we take all important lattice
configurations into account, we perform the integration in
Eq.(\ref{parfun}) over $\Delta(x)$ close to all possible
multikink configurations.  Thus, we first perform the
saddle-point integration near the configuration $\Delta_N(x |{\bf
z})$ with fixed number and positions of kinks, and then we
integrate over the positions $z_1 < z_2 < \ldots < z_N$ of
the kinks and sum over their number $N$.

The details of the saddle-point calculation are contained in the
Appendix. The resulting expression for the free energy of a chain
with disorder realization $\eta(x)$ is:
\begin{equation}
F[\eta(x)] = F_0 + F_1[\eta(x)]\;,
\label{F}
\end{equation}
where the first term is the free energy of a chain without kinks
and at zero disorder, and the second term is given by
\begin{equation}
F_1[\eta(x)] = - \frac{1}{\beta}
\ln \left[ \sum_N \frac{1}{d(T)^N}\int\!\!d^Nz
\exp \left\{ - \beta \left( \mu N -
\frac{2}{\pi \lambda v_F}
\int\!\! dx \Delta_N(x|{\bf z}) \eta(x)
\right) \right\} \right]\;,
\label{F1}
\end{equation}
Here $d(T)$ is a variable having dimension of length
defined by Eq.(\ref{d}).

We assume the size of the kinks to be much smaller than the
average distance between them.  In this case the kinks can be
replaced by abrupt steps,
\begin{equation}
\Delta_N(x|{\bf z }) \rightarrow \Delta_0 f_N(x|{\bf z}) \equiv
\Delta_0 \prod_{n = 1}^{N} \mbox{sign}(z_n - x)\;.
\label{f}
\end{equation}
Equation (\ref{F1}) can then be rewritten as
\begin{equation}
F_1[\eta(x)] = - \frac{1}{\beta}
\ln \left[ \sum_N \frac{1}{d(T)^N}\int\!\!d^Nz
\exp \left\{ - \beta \left( \mu N -
\kappa \int\!\! dx f_N(x|{\bf z}) \eta(x)
\right) \right\} \right]\;,
\label{FS1}
\end{equation}
where we introduced
\begin{equation}
\kappa = \frac{2 \Delta_0}{\pi \lambda v_F}\;.
\end{equation}

In the absence of disorder ($\eta(x) = 0$) the integration over
the kink positions in Eq.(\ref{FS1}) is trivial and the free
energy can be easily found,
\begin{equation}
F_1[\eta(x)=0] =  - \frac{L}{d(T) \beta} e^{- \beta \mu}\;.
\end{equation}
In this case there are only thermally-induced kinks and their
density,
\begin{equation}
\label{therm}
n_{therm} = \frac{1}{L} \frac{\partial F}{\partial \mu} =
\frac{1}{d(T)} e^{- \beta \mu}\;,
\end{equation}
is exponentially small at $T = \beta^{-1} \ll \mu$.  This
condition, which was already assumed to hold in writing
Eq.(\ref{adappr}), is always met for solitons in {\em
trans}-polyacetylene ($\mu \sim 0.5 eV$).

Disorder breaks the translational invariance.  Nevertheless, the
integration over the positions of kinks and the sum over their
number, can be performed immediately, by noting that
Eq.(\ref{FS1}) can be written in terms of the matrix elements of
an ordered exponential,
\begin{equation}
\label{ordexp}
F_1[\eta(x)] =  - \frac{1}{\beta}
\ln \sum_{\nu,\nu^{\prime}} \langle \nu^{\prime} |\,
{\mbox T} \exp \left[ \int_0^L\!\!dx
\left( \zeta(x) \sigma_3  + \rho \sigma_1
\right) \right]\!\!| \nu \rangle\;,
\end{equation}
where $\zeta(x) = \beta \kappa \eta(x)$, $\rho = e^{-\beta
\mu}/d(T)$, and $| \nu \rangle$ is an eigenvector of the
Pauli matrix $\sigma_3$, \[
\sigma_3 | \nu \rangle = \nu | \nu \rangle\;.
\]
The sum in Eq.(\ref{ordexp}) over $\nu,\nu^{\prime} = \pm 1$
corresponds to the sum over all possible signs of the
dimerization at the chain ends.  We stress that this choice of
boundary conditions is a matter of convenience and does not
affect our final result for the density of solitons.  The ordered
exponential describes the evolution (in imaginary time $x$) of a
spin $1/2$ in a magnetic field, which has a constant
$x$-component and a random time-dependent $z$-component.  The
corresponding time-dependent Hamiltonian is,
\begin{equation}
H(x) = \zeta(x) \sigma_3 + \rho \sigma_1\;.
\end{equation}
Kinks induced in a disordered chain correspond to spin flips and
the sum over all possible numbers of kinks and the integration
over their positions in Eq.(\ref{FS1}) correspond to the
expansion of the evolution operator in Eq.(\ref{ordexp}) in
powers of the random magnetic field.

It is now convenient to rotate the ``coordinate'' frame by an
angle $\pi/2$ around ${\bf e}_2$, which transforms the
Hamiltonian $H$ into
\begin{equation}
\label{transform}
H^{\prime}(x) =
\exp(i\frac{\pi}{4}\sigma_2)H\exp(-i\frac{\pi}{4}\sigma_2)
= -\zeta(x) \sigma_1 + \rho \sigma_3\;.
\end{equation}
The expression for the free energy then becomes,
\begin{equation}
\label{ordexpr}
F_1[\eta(x)] = - \frac{1}{\beta}
\ln \left(2 \langle + |\,
{\mbox T} \exp \left[\int_0^L\!\!dx H^{\prime}(x)
\right]\!\!| + \rangle \right) =
-\frac{1}{\beta} \ln \left(2 \psi_{\uparrow}(L)\right)\;,
\end{equation}
where we introduced the wave function of the spin,
$
\psi(x) = \left( \begin{array}{c}
\psi_{\uparrow}(x) \\
\psi_{\downarrow}(x)
\end{array} \right)
$,
that satisfies the equation,
\begin{equation}
\label{Schroe}
\frac{d\psi}{dx} = H^{\prime}(x) \psi\;,
\end{equation}
with the initial condition
$
\psi(0) = \left( \begin{array}{c}
1 \\ 0
\end{array} \right)
$.

We now introduce $v(x)$ and $\phi(x)$, such that
\begin{equation}
\psi = v \left( \begin{array}{c}
\cosh\left(\frac{\phi}{2}\right) \\ \\
\sinh\left(\frac{\phi}{2}\right)
\end{array} \right)\;.
\end{equation}
The equation for $v(x)$,
\begin{equation}
\frac{1}{v}\frac{dv}{dx} = \rho \cosh(\phi)\;,
\end{equation}
can be easily integrated, yielding
\begin{equation}
\label{psiup}
\psi_{\uparrow}(L) = \cosh\left(\frac{\phi(L)}{2}\right)
\exp \left[ \rho
\int_0^L\!\!dx \cosh(\phi(x))
\right]\;.
\end{equation}
The equation for $\phi(x)$ has the form of a Langevin
equation,\cite{Parisi}
\begin{equation}
\label{Langevin}
\frac{d\phi}{dx} = - \frac{dU(\phi)}{d\phi} + f(x)\;,
\end{equation}
with the potential,
\begin{equation}
\label{pot}
U(\phi) = 2 \rho \cosh(\phi)\;,
\end{equation}
and the random force,
\begin{equation}
f(x) = - 2 \zeta(x) = 2 \beta \kappa \eta(x)\;.
\end{equation}

Since the potential $U(\phi)$ grows at $\phi \rightarrow \pm
\infty$, the distribution of $\phi$ at large $x$ tends to the
equilibrium distribution, given by the Boltzmann formula,
\begin{equation}
\label{Peq}
P_{eq}(\phi) = e^{-\frac{U(\phi)}{T_f}} \Bigg /
\int_{-\infty}^{+\infty}\!\!d\phi
e^{- \frac{U(\phi)}{T_f}}\;.
\end{equation}
where the effective temperature, $T_f = 2 A (\beta \kappa)^2$, is
determined by the correlation function of the random force,
\begin{equation}
\langle f(x) f(y) \rangle = 4 A (\beta \kappa)^2 \delta(x-y)
\equiv 2 T_f \delta(x-y)\;.
\end{equation}

Assuming the chain length $L$ to be much larger than the average
distance between kinks, we can use the equilibrium distribution
function Eq.(\ref{Peq}) to average the free energy over the
disorder.  Using Eqs.  (\ref{ordexpr}) and (\ref{psiup}) and
neglecting $1/L$ terms, we obtain,
\begin{equation}
\label{ansf}
\langle F \rangle = F_0 - \frac{L\rho}{\beta}
\langle \cosh(\phi) \rangle_{eq}\;.
\end{equation}
The average of $\cosh(\phi)$ over the equilibrium distribution
function Eq.(\ref{Peq}) can be expressed in terms of the modified
Bessel functions,
\begin{equation}
\label{ch}
\langle \cosh(\phi) \rangle_{eq} =
\int_{-\infty}^{+\infty}\!\! d\phi P_{eq}(\phi) \cosh(\phi) =
\frac{K_1(z)} {K_0(z)}\;,
\end{equation}
with
$
z =  2\rho / T_f
\propto \exp \left( - \beta \mu \right).
$
As we assumed before already that $T \ll \mu$, we can use the
approximate expressions for the modified Bessel functions at
small value of the argument $z$.  The average free energy of a
disordered Peierls chain can then be written in the form,
\begin{equation}
\label{smallT}
\langle F \rangle = F_0
- L A \left(\frac{2 \Delta_0}{\pi \lambda v_F}\right)^2
\frac{1}{\mu + \frac{3}{2} T \ln\left(\frac{eT_0}{T}\right)}\;,
\end{equation}
where the temperature $T_0$ is defined by,
\begin{equation}
\label{T_0}
T_0 = \frac{1}{e \lambda v_F}
\left( \frac{4 A \Delta_0^2}{\pi \gamma \sqrt{cr}}
\right)^{\frac23}\;,
\end{equation}
and $\gamma = 1.781072...$, is the exponential of Euler's
constant.

\section{Average density of kinks}\label{density}

We now turn to the average density $n$ of disorder-induced kinks,
which is obtained by differentiating the free energy
Eq.(\ref{smallT}) with respect to $\mu$ (cf.  Eq.(\ref{therm})).
This leads to:
\begin{equation}
\label{densful}
n = \frac{A}{\left(\gamma_k \lambda v_F\right)^2}
\frac{1}{\left( 1 + \frac{3T}{2\mu}
\ln\left(\frac{eT_0}{T}\right)\right)^2}\;,
\end{equation}
which at zero temperature reduces to
\begin{equation}
\label{answer1}
n(T=0) = \frac{A}{\left(\gamma_k \lambda v_F\right)^2}\;.
\end{equation} This zero-temperature result was derived
in Ref.[\CITE{MFK}] by using scaling arguments.

Strictly speaking, however, we cannot put $T = 0$ in our general
result, because for $T < T_0$ the entropy of kinks,
\begin{equation}
\langle S_{kink} \rangle = -
\frac{\partial \langle F \rangle}{\partial T}
\propto \ln\left(\frac{T}{T_0}\right)\;,
\end{equation}
becomes negative and the density of kinks increases with
decreasing temperature.  We note that in the expression for the
free energy Eq.(\ref{smallT}), the kink energy $\mu$ is
effectively replaced by,
\begin{equation}
\label{muT}
\mu(T) = \mu + \frac{3}{2} T \ln \left( \frac{e T_0}{T}
\right)\;.
\end{equation}
This can be readily understood as follows. Each kink enters in
the expression Eq.(\ref{F}) with a weight:
\begin{equation}
\label{muT2}
e^{-\beta \mu(T)} = \frac{l(T)}{d(T)} e^{-\beta \mu}\;,
\end{equation}
where $l(T)$ is the typical size of the thermal fluctuation of
the kink position. For the white noise disorder that we consider,
$l(T)$ can be estimated from the condition,
\begin{equation}
(\beta \kappa)^2 A l(T) \sim 1\;.
\label{1l(T)}
\end{equation}
This derives from the criterion that shifting the kink over
$l(T)$ from its optimal position, should cause a fluctuation of
the order of $T$ in its interaction energy Eq.(\ref{EN}) with the
disorder.  Using Eq.(\ref{d}) we find that,
\begin{equation}
\frac{l(T)}{d(T)} \propto T^{\frac32}\;,
\label{2l(T)}
\end{equation}
which in combination with Eq.(\ref{muT2}) indeed gives
Eq.(\ref{muT}) for $\mu(T)$.  The temperature dependence of
$\mu(T)$ is plotted in Fig.~2.  Below $T = T_0$,
$\frac{d\mu(T)}{dT} > 0$, {\em i.e.}, kinks become less
``heavy'' as temperature decreases, which explains the
pathological behavior of the entropy.

Keeping in mind that our description of the lattice is both
classical and continuum, this pathology is hardly suprising.  In
practice, quantum effects will prevent the entropy from becoming
negative at low temperatures.  A simpler way to regularize the
model, lies in a discretization.  From Eqs.(\ref{muT}) and
(\ref{muT2}), it follows that $T_0$ is determined by the
condition $l(T) \sim d(T)$.  For smaller temperatures, $l(T)$
gets smaller than $d(T)$, which plays the role of an effective
lattice constant and becomes arbitrarily small at small
temperatures.  Clearly, however, the continuum description of the
off-diagonal disorder fails when $l(T)$ drops below the lattice
constant $a$.  Moreover, the Peierls-Nabarro
barrier\cite{Nabarro} at low temperature fixes the kink position
within the unit cell of the atomic lattice.  These facts motivate
us to consider in the next section a discrete version of
Eq.(\ref{F}), in which a temperature-independent lattice constant
$d$ is used and the integrations over kink positions are
substituted by finite lattice sums.

It should be noted that just replacing $d(T)$ in Eq.(\ref{F}) by
a temperature-independent distance $d$ does not, by itself, make
the entropy positive.  Retracing the steps following Eq.(\ref{F})
in that case leads to
\begin{equation}
\label{Ffixd}
\langle F \rangle = F_0
- L A \left(\frac{2 \Delta_0}{\pi \lambda v_F}\right)^2
\frac{1}{\mu + 2T \ln\left(\frac{eT_{0}^\prime}{T}\right)}\;,
\end{equation}
with
\begin{equation}
\label{T_0pr}
T_{0}^\prime = \frac{2\Delta_0}{\pi e \lambda v_F}
\left( \frac{2 d A}{\gamma} \right)^{\frac12}\;.
\end{equation}
The density of disorder-induced kinks is then found to be
\begin{equation}
\label{densful2}
n = \frac{A}{\left(\gamma_k \lambda v_F\right)^2}
\frac{1}{\left( 1 + \frac{2T}{\mu}
\ln\left(\frac{eT_{0}^\prime}{T}\right)\right)^2}\;.
\end{equation}

Obviously, this result has the same zero-temperature limit as
Eq.(\ref{densful}) and it also shows a similar pathology below a
certain temperature, which now is given by $T_{0}^\prime$.  Only
if we discretize the kink positions on the lattice with constant
$d$, we will indeed obtain a classical model that also at low
temperatures gives sensible results.

\section{Discrete model and numerical results}\label{num}

As motivated in the previous section, we now consider a discrete
model, in which the chain is divided into $M$ cells of length $d$
(independent of $T$) and kinks are allowed to be located only at
the boundaries $x_m = m d$ ($m =1, \ldots, M$) between the cells.
Then the function $f(x) = \frac{\Delta(x)}{\Delta_0}$ [cf.
Eq.(\ref{f})] has a constant value $\pm 1$ inside each cell.  In
other words, the chain configuration is described by a set of
Ising variables, $\{ \sigma_1,\sigma_2,\ldots,\sigma_M\}$, where
$\sigma_m$ is the value of $f(x)$ in the $m$th cell.  We impose
periodic boundary conditions $\sigma_{M+1} = \sigma_1$.  Kinks
occur between cells with opposite values of $\sigma$ and the sum
over all possible configurations of the Ising spins represents
the sum over the number of kinks, as well as over their positions
on the one-dimensional lattice.  This lattice model for the
statistics of disorder-induced kinks is equivalent to the
one-dimensional Random Field Ising Model\cite{IM} (RFIM).  The
constant-$d$ continuum model which led to Eq.(\ref{densful2}) is,
in fact, the continuum version of this RFIM.

The partition function of the one-dimensional RFIM reads
\begin{equation}
Z\;=\;\sum_{\{\sigma_{m}\}}\,
e^{-\beta E\{\sigma_{m}\}}\;,
\label{zrfim}
\end{equation}
with
\begin{equation}
E\{\sigma_{m}\}\;=\sum_{m=1}^{M} \left[
\mu \frac{\left(1-\sigma_{m}\sigma_{m+1}\right)}{2}\;
-\;h_{m}\sigma_{m}\,\right]\;,
\label{erfim}
\end{equation}
The first term in Eq.(\ref{erfim}) equals $\mu$
times the number of spin flips in the chain and thus represents
the total kink creation energy.  The second term describes the
interaction with the random ``magnetic'' field defined by,
\begin{equation}
h_{m}\;=\;\kappa\,\int_{x_m}^{x_{m+1}}\!\!dx\eta(x)\;,
\label{randfield}
\end{equation}
{\em i.e.}, the value of the field in each cell is proportional
to the value of disorder in the continuum model integrated over
the cell.  The distribution of the random field is also
Gaussian, with correlator:
\begin{equation}
\langle h_k h_l \rangle = D \delta_{kl}\;,
\end{equation}
where $D = d \kappa^2 A$.

It is well-known that arbitrarily small disorder or temperature
destroy the long range order in the one-dimensional
RFIM.\cite{IM} The density of the thermally excited domain walls
in the absence of disorder is $e^{- \beta \mu}$, while at zero
temperature the density of the disorder-induced kinks is
proportional to the strength of the disorder.  A full calculation
of the average free energy and the density of kinks in the RFIM
involves numerical simulations.  We performed such simulations
using the transfer-matrix approach, as we did in the continuum
treatment.  In the discrete case, the Schr\"odinger equation
(\ref{Schroe}) for the continuum model is replaced by
\begin{equation}
\psi_{m+1} = {\hat{T}_{m}} \psi_m\;,
\end{equation}
with the transfer matrix $\hat{T}_{m}$,
\begin{equation}
\hat{T}_{m}\;=\;
\left(\begin{array}{cc}
\exp\{\beta h_{m}\} & \exp\{-\beta (\mu-h_{m})\} \\ \\
\exp\{-\beta (\mu+h_{m})\} & \exp\{-\beta h_{m}\}
\end{array} \right)\;,
\label{tmatrix}
\end{equation}
The free energy for a given realization $\{ h_m \}$ of the random
field is then given by
\begin{equation}
F[\{h_m\}] = - T \ln \left( \mbox {Tr} \prod_{m=1}^M {\hat T}_m
\right)\;.
\end{equation}
To obtain a smooth temperature dependence of the density of
kinks, we had to average the free energy over $10^3$ random field
realizations for a chain with $10^3$ sites.

We now turn to the results of these simulations.  In Fig.~3 the
free energy is plotted as a function of temperature for two
values of the disorder strength: $D = 0.005\mu^{2}$ and $D =
0.01\mu^{2}$.  Stars indicate the numerical results for the
discrete model, while the solid lines represent the continuum
version of the RFIM (Eq.(\ref{Ffixd})).  The latter curves have
maxima at $T=T_{0}^\prime$, below which the entropy becomes
negative.  For convenience, we also plotted the dashed curve,
which passes through the maxima of the solid lines.  The
continuum results become meaningless to the left of this curve.
The slope of the numerically obtained free energy of the discrete
model is always negative, corresponding to a positive entropy,
which tends to $0$ as $T \rightarrow 0$.  We also observe that
above $T_{0}^\prime$ the numerical (discrete) and analytical
(continuum) results only differ little from each other.

In Fig.~4 we give the density of kinks as a function of
temperature for the same two values of the disorder as in Fig.~3.
Again, solid lines and stars represent the continuum model and
the discrete model, respectively, while the dashed curve passes
through the minima of the continuum results, below which the
continuum model becomes meaningless.  The temperature dependence
of the density of kinks for $T < 0.15 \mu$ is very slow.  At $T =
0.15\mu$ the density of thermally induced kinks in a chain
without disorder is about $1.3 \cdot 10^{-3}/d$.  This means that
the increase of the kink density observed in Fig.~4 is mostly
related to the fact that more multikink configurations become
available in a disordered chain as temperature grows.

Figure 5 shows the density of kinks as a function of the disorder
strength $D$ at $T=0$.  Stars give the numerical results for the
discrete model (simulated at $T = 0.005 \mu$ in order to avoid
numerical instabilities).  The dashed line gives the $T=0$
continuum result, Eq.(\ref{answer1}), which in terms of $D$ reads
$n(T=0)=D/(d\mu^2)$.  As is observed from Fig.~4, for decreasing
disorder these two results get in better agreement with each
other.  This is a consequence of the fact that $T_{0}^\prime$
then also decreases.  In the limit of $D \rightarrow 0$, the
agreement should become exact, as indeed appears from Fig.~5.
For higher disorder, however, the nonphysical nature of the
continuum model for $T<T_{0}^\prime$ causes the numerical and
analytical results to deviate by an increasing amount at $T=0$.
An alternative approach to obtain an analytical zero temperature
result, is to use the continuum result Eq.(\ref{densful2}) at the
lowest possible value where the result is still physical, {\em
i.e.}, at $T=T_{0}^\prime$.  The thus obtained value of the
density of kinks as a function of the disorder is also plotted in
Fig.~5 (solid line) and is indeed seen to be in much better
agreement with the numerical $T=0$ results over a large interval
of the disorder.  This shows the importance of the corrections to
the linear $D$ dependence, expected on the basis of scaling
arguments,\cite{IM} of the density of domain walls in the RFIM.

\section{Summary and discussion}\label{dis}

In this paper we studied a continuum model describing Peierls
chains with a doubly degenerate ground state in the presence of
weak off-diagonal disorder.  In such chains, thermally- and
disorder-induced neutral solitons (dimerization kinks) occur.  By
integrating out all irrelevant lattice degrees of freedom, we
obtained the free energy of these kinks.  This free energy was
then averaged over the disorder realizations leading to an
analytical expression Eq.(\ref{densful}) for the average density
of kinks.

In the continuum model, the entropy of the kinks was found to
become negative below a certain temperature $T_0$.  At this
temperature the thermal fluctuation of the kink positions $l(T)$
becomes of the order of the effective ``lattice constant''
$d(T)$.  It was then argued that below this temperature one
should either take the effects of the lattice discreteness into
account, or consider the lattice quantum-mechanically.  We noted
that the discrete version of the model that describes the thermal
or disorder-induced kinks, is the one-dimensional RFIM.  The free
energy and the density of kinks in the latter model were found
numerically and the results were compared with the continuum
model.  It was found that for $T>T_0$, both models are in good
agreement.

To obtain the free energy of the kinks (Appendix A), we used
a technique developed for the semiclassical solution of quantum
problems.\cite{Coleman} This is not accidental, because a Peierls
chain with a doubly degenerate ground state in the presence of
weak disorder or at a small temperature behaves similarily to a
quantum particle in a double well.  If the wells are very deep,
the small quantum fluctuations around the two classical vacua may
be relatively insignificant.  A more important effect, however,
is that the quantum particle can tunnel between the two minima.
The tunneling probability is described by instantons, which are
the imaginary-time classical trajectories leading from one
minimum of the potential to the other.  In fact, an exact mapping
exists between the instanton gas for a double-well potential and
the gas of thermally-induced kinks.\cite{Polyakov} The
disorder-induced kinks are somewhat different: their positions,
unlike the instanton positions, are not arbitrary.  These kinks
are, to some extent, pinned by the disorder which induces them.

The main assumption of our calculation is the validity of the
expansion of the free energy in powers of the disorder.  As was
already discussed in Ref.[\CITE{MFK}], this translates into the
condition:
\begin{equation}
\label{condition1}
n \xi_0 \ll 1\;,
\end{equation}
meaning that the average distance between kinks should be much
larger than their size.  For stronger disorder the lattice
distortions would be random over the length scale of
the order of $\xi_0$, and the notion of multikink configurations
would lose sense.

Effects of disorder on the properties of Peierls chains have been
studied before by a number of authors.  Many of these studies
were devoted to charged solitons induced by random doping\cite
{dop} and the effects of localized bond and site impurities
on the electronic states.\cite{dis} The minimal-energy lattice
configuration in the presence of a finite density of bond
impurities was studied numerically in Refs.~[\CITE{Iw,Wa}].
There it was found that for some disorder realizations the
configuration contained kinks.  As these calculations were
performed within the SSH model (in which electron-electron
interactions are not included), the obtained disorder-induced
kinks did not have a definite charge or spin.  Also, no attempt
was made in these papers to obtain (numerically or analytically)
the density of solitons as a function of the disorder strength.

Another large body of work on disordered Peierls systems deals
with the Fluctuating Gap Model (FGM).\cite{Keldysh} This work not
only addresses the study of disorder {\em per se},\cite{OE,XT}
but also includes studies of quantum lattice fluctuations modeled
by static disorder.\cite{Kim,HM,MS} In the FGM, one assumes that
the lattice part $\Delta_{lat}$ of the order parameter is a
constant throughout the chain and does not respond to the local
disorder.  This immediately eliminates the possibility of
disorder-induced solitons.  Nevertheless, solitons play an
important role in the FGM: we recently showed that the optimal
fluctuation that determines the density of states inside the
pseudogap in this model has the form of a soliton-antisoliton
pair.\cite{MK2}

We finally mention that, while we considered weak off-diagonal
disorder, one may also consider a type of disorder, where the
electron conjugation over certain bonds is broken completely
($t_{n,n+1}=0$).  This model leads to a distribution of chain
sizes.\cite{Silbey} If the conjugation breaks occur randomly,
chains with an odd number of carbon atoms will occur, which
necessarily contain a kink.\cite{SU} Thus, the density of kinks
is proportional to the disorder strength ({\em i.e.} the density
of chain breaks), as it is in our model (see Eq.(\ref{answer1})).

As was already discussed in our previous paper,\cite{MFK} the
electron state in the presence of neutral solitons is different
from that of a perfectly dimerized chain.  This gives the
possibility to detect the disorder-induced kinks.  First, the
kinks should give rise to a peak in the optical absorption
spectrum that occurs inside the gap of the perfectly dimerized
chain.  Second, having spin $1/2$, the neutral kinks contribute
to the Curie susceptibility.  As we noted in Ref.[\CITE{MFK}],
experiments on undoped {\em trans}-polyacetylene do not seem to
agree with the large density of free solitons that one would
expect on the basis of the small average electron conjugation
length (several tens of carbon atoms\cite{Silbey}) in this
material.  In particular, experiments indicate that only 1 free
spin occurs per 3000 carbon atoms.\cite{SII} It should be noted,
however, that a meaningful comparison between theory and
experiment can only be made if one accounts for interchain
interactions, which suppress the generation of isolated
kinks.\cite{BM} A study of the interplay between the
disorder-induced kinks, destroying the long range coherence, and
the interchain interactions, binding kinks into pairs, will be
reported elsewhere.  It should be noted that this interplay seems
to be very important to describe the Peierls phase
transition.\cite{KGM}

We finally note that another important issue is the quantum
treatment of the lattice dynamics coupled to the spin dynamics of
the kinks.  The interaction between neighboring kinks may bind
their spins into a singlet ground state, thus explaining why the
density of free spins observed in undoped {\em
trans}-polyacetylene is so low.  We plan to address this problem
in the future.

\section*{Acknowledgments}

This work is supported by the "Stichting Scheikundig Onderzoek in
Nederland (SON)" and the "Stichting voor Fundamenteel Onderzoek
der Materie (FOM)".

\appendix
\section{}
\label{part}

In this Appendix we obtain the effective free energy of kinks by
integrating out all irrelevant lattice degrees of freedom in the
partition function of the disordered chain Eq.(\ref{parfun}).

It is convenient to replace the integration over lattice
configurations $\Delta_{lat}$ by the integration over
$\Delta(x)$, which also includes the disorder [cf.
Eq.(\ref{odpar})] .  The chain energy can be written as,
\begin{equation}
E[\Delta_{lat}(x),\eta(x)] = E_{el}[\Delta(x)] +
E_{lat}[\Delta(x) - \eta(x)]\;\;.
\end{equation}
Since the lattice energy is quadratic in $\Delta_{lat} =
\Delta(x) - \eta(x)$, we obtain,
\begin{equation}
E[\Delta_{lat}(x),\eta(x)] = E[\Delta(x),0] -
\frac{2}{\pi \lambda v_F} \int_0^L\!\!dx \Delta(x) \eta(x) +
\frac{1}{\pi \lambda v_F} \int_0^L\!\!dx \eta(x)^2\;\;,
\end{equation}
where the first term is the chain energy at zero disorder and $L$
is the chain length.

For $\Delta(x)$ close to the multikink configuration $\Delta_N(x
|{\bf z})$ we can expand the chain energy $E[\Delta(x),0]$ in
powers of $\xi(x)$ defined by:
\begin{equation}
\xi(x) = \Delta(x) - \Delta_N(x |{\bf z})\;.
\end{equation}
Retaining terms up to second order in $\xi(x)$, we have
\begin{equation}
\label{expan}
E[\Delta,0] = E_0 + N \mu +
\frac{1}{\pi \lambda v_F} \int_0^L\!\!dx
\xi(x) {\hat D}_N({\bf z}) \xi(x)\;.
\end{equation}
Here, we assumed the distances between kinks to be much larger
than the correlation length $\xi_0$.  As we discussed in
Sec.~\ref{stabil}, in the absence of disorder the multikink
configuration is then an almost exact extremum of the chain
energy, which is why Eq.(\ref{expan}) does not contain a
first-order term.  The operator ${\hat D}_N({\bf z})$ is defined
through:
\begin{equation}
{\hat D}_N({\bf z})\xi(x)
= \int\!\!dy D_N(x,y|{\bf z}) \xi(y)\;,
\end{equation}
with the kernel $D_N(x,y|{\bf z})$ related to the second
variational derivative of the total chain energy:
\begin{equation}
\frac{1}{\pi \lambda v_F}
D_N(x_1,x_2|{\bf z}) =
\frac{1}{2} \left.
\frac{\delta^2 E [\Delta,0]}{\delta\Delta(x_1) \delta\Delta(x_2)}
\right|_{\Delta = \Delta_N(x |{\bf z})}\;\;.
\end{equation}
Let $\Omega = \sqrt{\frac{4K}{M}}$ denote the bare optical phonon
frequency.  Then, the eigenfunctions $\phi_{\alpha}(x|{\bf z})$
of ${\hat D}_N({\bf z})$ obeying,
\begin{equation}
{\hat D}_N({\bf z})\phi_{\alpha}(x|{\bf z}) =
\left( \frac{\omega_{\alpha}({\bf z})}{\Omega} \right)^2
\phi_{\alpha}(x|{\bf z})\;,
\end{equation}
are (optical) phonon modes in the presence of $N$ kinks with
frequencies $\omega_{\alpha}({\bf z})$.

When the kinks are far apart, the $N$ lowest phonon frequencies
($\alpha = 1,\ldots,N$) are very small, since these normal modes
correspond to shifts of the kinks, which leave the chain energy
practically unchanged.  (The frequency of one mode is exactly
zero, as it corresponds to the translation of the multikink
configuration as a whole).  To avoid double counting, the
integration over the first $N$ modes is replaced by the
integration over the $N$ kink positions, which is the standard
approach in the instanton calculus.\cite{Coleman}  Then the
measure of the functional integration in the vicinity of the
multikink configuration becomes,
\begin{equation}
D\xi(x) = J_N d^Nz \prod_{\alpha > N} d\xi_{\alpha}\;.
\end{equation}
where $\xi_{\alpha}$ are the amplitudes of the normal modes,
\begin{equation}
\label{expans}
\xi(x) = \sum_{\alpha} \xi_{\alpha}
\phi_{\alpha}(x|{\bf z})\;,
\end{equation}
and $J_N$ is the Jacobian of the substitution of the integration
over the $N$ (almost) zero modes by the integration over the kink
positions.  We have:
\begin{equation}
\label{Jac}
J_N = (J_1)^N \equiv c^{\frac{N}{2}}\;,
\end{equation}
with $J_1$ the Jacobian for the single-kink configuration
$\Delta_1(x)$, leading to
\begin{equation}
c = \int\!\!dx
\left( \frac{d \Delta_1(x)}{dx} \right)^2
\sim \frac{\Delta_0^2}{\xi_0}\;.
\label{c}
\end{equation}
The constant $c$ can be expressed as $c = \pi \lambda v_F
\Omega^2 M_k/2$, with $M_k$ the soliton mass, which in the SSH
model is about $6 m_e$.\cite{SSH}

The Gaussian integration over the amplitudes of the non-zero
modes now reduces the partition function to,
\begin{equation}
Z = Z_0 \sum_N
\left(\frac{4 \beta c}{\pi^2 \lambda v_F} \right)^{\frac{N}{2}}
\int\!\!d^Nz R_N({\bf z})
\exp \left \{ - \beta W_N({\bf z}) \right \}\;.
\label{Z}
\end{equation}
Here
\begin{equation}
W_N({\bf z}) =  N \mu -
\frac{2}{\pi \lambda v_F} \int_0^L\!\!dx \Delta_N(x|{\bf z})
\eta(x) + \frac{1}{\pi \lambda v_F} \left(
\int_0^L\!\!dx \eta(x)^2 -
\sum_{\alpha > N}
\frac{\Omega^2}{\omega_{\alpha}^2({\bf z})}\eta_{\alpha}^2
\right)\;,
\label{W_N}
\end{equation}
with $\eta_{\alpha}$ the expansion coefficients of $\eta(x)$ over
the orthonormal basis $\phi_{\alpha}(x|{\bf z})$ (cf.
Eq.(\ref{expans})).  $Z_0$ is the partition function of the chain
in the absence of kinks and disorder:
\begin{equation}
Z_0 = e^{-\beta E_0}
\prod_{\alpha} \left(
\pi \sqrt{\frac{\lambda v_F}{\beta}}
\frac{\Omega}{\omega_{\alpha}(0)} \right)\;,
\end{equation}
where $\omega_{\alpha}(0)$ denote the phonon frequencies in the
absence of kinks and disorder.  Finally, the factor $R_N({\bf
z})$ in Eq.(\ref{Z}) is the product of nonzero phonon frequencies
for the lattice configuration containing no kinks divided by
the same product for the configuration with $N$ kinks,
\begin{equation}
R_N({\bf z}) =
\left( \prod_{\alpha}
\frac{\omega_{\alpha}(0)}{\Omega} \right)
\left(
{\prod_{\alpha > N}\frac{\omega_{\alpha}({\bf z})}{\Omega}}
 \right)^{-1}\;.
\end{equation}
Similarily to Eq.(\ref{Jac}), for widely separated kinks,
$R_N({\bf z})$ equals the $N$th power of the regularized product
of nonzero phonon frequencies for a configuration with one
kink,\cite{Coleman}
\begin{equation}
R_N({\bf z}) = \left(R_1(z_1)\right)^N \equiv
r^{\frac{N}{2}}\;,
\label{Rr}
\end{equation}
where $r$ is a number of the order unity.

Finally, for weak disorder we can neglect the third term in
Eq.(\ref{W_N}), as it is of second order in $\eta(x)$. From
Eqs.(\ref{Z}), (\ref{W_N}), and (\ref{Rr}) we then obtain the
expressions Eqs.(\ref{F}) and (\ref{F1}) for the free energy of a
chain with disorder realization $\eta(x)$, where $d(T)$ is given
by
\begin{equation}
d(T) = \frac{\pi}{2} \sqrt{\frac{\lambda v_F T}{c r}}
\sim \xi_0 \sqrt{\lambda \frac{T}{\Delta_0}}\;.
\label{d}
\end{equation}

In this calculation we neglected electron excitations, assuming
the temperature to be sufficiently low.  However, the electron
ground state for the lattice configuration containing $N$ neutral
kinks, separated by distances much larger than the correlation
length, is $2^{N}$-fold degenerate, due to the fact that the spin
projection of each kink can be arbitrary.  Because of this
degeneracy, the weight factor of each kink acquires an extra
factor of $2$, which was taken into account in Eq.(\ref{Z}).  The
degeneracy is not exact: excitations of this spin system have a
typical energy of the order of the exchange constant for the
spins of two neighboring kinks.  Thus, our assumption requires
that the temperature is much larger than this energy [$\Delta_0
\exp(-1/(n\xi_0))$], yet much smaller than the Peierls gap
[$\Delta_0$].

\newpage

\begin{center}
\large
{\bf  FIGURE CAPTIONS}
\end{center}

\vspace{2cm}

FIG.1   Order parameter along a Peierls chain for a multikink
configuration with 7 kinks. $z_1, z_2, \ldots, z_7$, denote the
kink positions.

FIG.2   Temperature dependence of the effective chemical
potential of kinks according to Eq.(\ref{muT}) for $\mu = 10
T_0$.

FIG.3.  The free energy of dimerization kinks in a Peierls chain
as a function of temperature for two values of the disorder
strength: $D = 0.005\mu^{2}$ (a) and $D = 0.01\mu^{2}$ (b).  The
solid lines give the analytical result Eq.(\ref{Ffixd}), while
the stars indicate the numerical results.  The part of the free
energy related to an ordered chain without kinks ($F_0$) is
discarded.  $M=L/d$ denotes the number of unit cells.  The dashed
curve intersects the solid lines at the temperature for which the
analytically calculated entropy equals zero.

FIG.4.  The density of dimerization kinks in a Peierls chain as a
function of temperature for two values of the disorder strength:
$D = 0.005\mu^{2}$ (a) and $D = 0.01\mu^{2}$ (b).  The solid
lines give the analytical result Eq.(\ref{densful2}) and the
stars indicate the numerical results.  Dashed curve as in Fig.~3.

FIG.5.  Density of dimerization kinks in a Peierls chain as a
function of disorder strength.  The dashed line represents the
analytical continuum result Eq.(\ref{densful2}) taken at $T=0$,
the solid line gives the same result at $T=T_{0}^\prime$, while
the stars indicate the results of numerical simulations of the
discrete model at $T = 0.005 \mu$ (which for all practical
purposes may be considered zero temperature).

\end{document}